\documentclass[ras]{agutex}
\usepackage[dvips]{graphicx}
\setkeys{Gin}{draft=false}

\authorrunninghead{ROGERS ET AL.}

\titlerunninghead{MEASUREMENTS OF ELECTRON TEMPERATURE}

\authoraddr{Corresponding author: A. E. E. Rogers,
M.I.T. Haystack Observatory
off route 40, Westford, Massachusetts 01886 USA.
(arogers@haystack.mit.edu)}

\begin{document}

\title{Radiometric Measurements of Electron Temperature and Opacity of Ionospheric Perturbations}

%Use \author{\altaffilmark{}} and \altaffiltext{}

% \altaffilmark will produce footnote;
% matching \altaffiltext will appear at bottom of page.

\authors{A. E. E.  Rogers\altaffilmark{1}
J. D. Bowman,\altaffilmark{2} 
J. Vierinen, \altaffilmark{1}
R. Monsalve,\altaffilmark{2}
T. Mozdzen\altaffilmark{2}}

\altaffiltext{1}{M.I.T. Haystack Observatory,
Westford, Massachusetts, USA.}

\altaffiltext{2}{Arizona State University, Tempe, Arizona, USA.}

\begin{abstract}
Changes in the sky noise spectrum are used to characterize perturbations in the ionosphere.   Observations were made at the same sidereal time on multiple days using a calibrated broadband dipole and radio spectrometer covering 80 to 185 MHz.  In this frequency range, an ionospheric opacity perturbation changes both the electron thermal emission from the ionosphere and the absorption of the sky noise background.  For the first time, these changes are confirmed to have the expected spectral signature and are used to derive the opacity and electron temperature associated with the perturbations as a function of local time. The observations were acquired at the Murchison Radio-astronomy Observatory in Western Australia from 18 April 2014 to 6 May 2014.  They show perturbations that increase at sunrise, continue during the day, and decline after sunset. Magnitudes corresponding to an opacity of about 1 percent at 150~MHz with a typical electron temperature of about 800~K, were measured for the strongest perturbations.
\end{abstract}

\begin{article}

\section{Introduction}

The ionospheric effects on radio propagation, which are fairly small in the
VHF frequency band during normal ionospheric conditions, are still
significant enough that they need to be taken into account for a ground
based measurement of the signature of 
the epoch of
reionization (EoR) [{\textit{Bowman and Rogers}}, 2010]. An overview of these
ionospheric chromatic effects is given by \textit{Vedantham et. al,} [2010].

In this study, observational results of ionospheric effects in the
80-185 MHz frequency range are presented, including the first
spectrometric detection of ionospheric radio emission, allowing the
measurement of electron temperature in addition to ionospheric
absorption. Understanding and being able to measure these effects is not
only important for future measurements of the spectral signature of the
EoR signature, but could also have applications in ionospheric remote
sensing. Currently electron temperatures in the ionosphere can be
measured either using incoherent scatter radars, or satellite-borne in
situ measurements with Langmuir probes.

A riometer [{\textit {Little and Leinbach}, 1959}] is an instrument that is used to 
measure ionospheric absorption. This instrument uses an antenna and a 
stable radio frontend to measure cosmic radio noise to infer the amount 
of absorption occurring in the ionosphere. The noise temperature measured 
by the instrument is equal to the sky noise brightness convolved with 
the antenna beam. This is a repeatable function of sidereal time if the 
variable effects of solar radiation, ionospheric absorption, and 
ionospheric emission are excluded. At frequencies below about 100 MHz, 
ionospheric absorption dominates. In this case observations of a 
decrease of the noise power from the maximum sky noise observed at that 
sidereal time can be used to measure the ionospheric absorption. The 
"reference" curve of maximum power as a function of sidereal time can be 
measured at the night when the ionosphere is at minimum and the radio 
noise from the Sun is excluded. The receiver used in a riometer needs to 
be very stable in time, although accurate absolute calibration is not 
required.

While typically a riometer uses only a single
frequency, the measurements reported in this paper were made with a
wideband instrument and cover 80 to 185~MHz.  This instrument was designed as part of the Experiment to Detect the Global EoR Signature (EDGES) 
for astronomical observations of the early universe [\textit{Bowman and Rogers}, 2010; \textit{Bowman et al.}, 2008; \textit{Rogers and Bowman}, 2008].  
Like a riometer, it is very stable and repeatable in time.  
Further, its frequency response is very well calibrated (see \textit{Rogers and Bowman} [2012] for details of the calibration method and the instrument) 
so that observed changes in the spectrum taken at the same time on successive
days can be used to measure changes in ionospheric emission as well as
absorption.

Observations reported here,
which were made using the EDGES instrument deployed at the Murchison Radio-astronomy Observatory (MRO) at $(-26.7^\circ N ~,~ 116.6^\circ E)$ 
in Western Australia from 18 April 2014 to 6 May 2014,
 may not be the first radiometric measurements of the electron temperature in the ionosphere. \textit{Pawsey et al.} [1951] 
report emission temperatures in the range 240 to 290 K 
at 1.6 and 2.0 MHz  which could have been from the lower ionosphere. 
However, while at these low frequencies the emission from sky is completely reflected back into space 
during daytime, thermal emission 
from the earth can be propagated via a ground wave or be 
reflected from the ionosphere and atmospherics from lightning as well as man-made radio frequency interference (RFI). This makes observations at these low frequencies very difficult.  
 A broadband spectrometer, such as the EDGES instrument, has the advantage over a traditional riometer that contaminating changes in the observed sky power due to solar emission, 
RFI, and the effects of severe weather conditions can be distinguished based on their spectral properties from the absorption and emission due to the ionosphere.

We describe our analysis method in Section 2.  In Section 3, we present the data and principal results of the analysis,
followed by interpretation and discussion in Section 4 and concluding remarks in Section 5.

\section{Method}

The noise power temperature from an antenna at a given frequency is equal to the
sky noise brightness, $T_{sky}(\vec\theta, \nu, t)$, convolved with the antenna beam
directivity, $D(\vec\theta, \nu)$,

\begin{equation}
T_{ant}(\nu, t) = T_{sky}(\vec\theta, \nu, t) \otimes D(\vec\theta, \nu)
\end{equation}
where $\vec\theta$ is sky coordinate, $\nu$ is frequency, and $t$ is time [{\textit {Rogers et al.}}, 2004].
In the absence of the ionosphere, RFI and solar flares the spectrum, $T_{ant}(\nu)$, is a highly repeatable function of sidereal time.
Its spectral shape follows the beam averaged power law of the sky brightness provided the beam shape is constant with frequency.

\subsection{ Ionosphere model}

If an opacity change occurs in a region of uniform electron
temperature, the change in the spectrum from the theory of radiative
transfer [{\textit {Chandrasekhar}, 1960}] is approximately

\begin{equation}
 (-T_Bf^{-s}+T_e)f^{-2}\Delta\tau
\end{equation}
where $T_B$ is $T_{ant}$ evaluated at 150~MHz, $T_e$ is the electron temperature of the perturbed region, $f$ is 
the normalized frequency given by $\nu/(150$~MHz), $\Delta\tau$ is the change in opacity at 150~MHz, 
and $s$ is the spectral index of the sky brightness power law spectrum. 

This approximation is valid for the small opacity and small changes in
the ionosphere expected in the frequency range 80 to 185~MHz.
The spectral index is approximately 2.5 in this frequency range  [{\textit {Rogers and Bowman}}, 2008]. 
Typical attenuation through the ionosphere at 150~MHz is 0.015~dB at night and
0.1~dB during the day and is inversely proportional to frequency
squared ({\textit {Evans and Hagfors}}, 1968 and references therein).

\subsection{Analysis}

Two methods are used to analyze difference spectra
for electron temperature and change in opacity.
In the first method the
average and standard deviation of the weighted electron temperature
and average magnitude of the perturbations can be estimated from the
difference spectra taken at the same sidereal time on different days.
If there are $N$ days of data there will be $N(N-1)/2$ difference
spectra that can be analyzed. Each difference spectrum is fitted to
the function

\begin{equation}
af^{-2-s} + bf^{-2}
\end{equation}
using weighted least squares. The perturbation in opacity and
associated electron temperature are then derived from the best fit
parameters $a$ and $b$, resulting in

\begin{equation}
\Delta\tau = -aT_B^{-1}
\end{equation}
and

\begin{equation}
     T_e = -(b/a)T_B.
\end{equation}

In order to improve the detection and measurement of weak
perturbations, their average magnitude and associated electron
temperature can be estimated by conducting incoherent averaging of the coefficients
derived from the difference spectra.  Squaring equation (4)  and
multiplying equation (5) by $a$ eliminates the division by $a$ in
equation (5) which results in

\begin{equation}
 (\Delta\tau)^2 = a^2 T_B^{-2}
\end{equation}
and

\begin{equation}
T_e = -abT_B / a^2.
\end{equation}

The bias in $a^2$ and $ab$ due to Gaussian noise in the spectra can be
removed by subtracting the noise expected in the absence of
perturbations,

\begin{equation}
\overline{a^2} = \ \langle a^2\ - C_{aa}\ \mathrm{rms}^2\rangle
\end{equation}
and
 
\begin{equation}
\overline{ab} = \ \langle ab\ -  C_{ab}\ \mathrm{rms}^2 \rangle
\end{equation}
where $C_{aa}$ and $C_{ab}$ are the values from the covariance matrix 
used to fit each difference spectrum
and $\mathrm{rms}$ is the root mean square (RMS) residual from the fit
used to obtain the values of $a$ and $b$. This method is similar to
the method of bias removal used for the incoherent averaging of
interferometer correlation amplitudes [\textit{Rogers, Doeleman and
    Moran}, 1995].

A sequence of 16~days of data from day 108~to 126 (excluding days 109,
115, 125) was analyzed in half hour integrations at the same sidereal
time each day. The 16 days provide up to 120 difference spectra. A
weighted average of the noise-corrected values of $\overline{a^2}$ and
$\overline{ab}$ is made to improve the signal-to-noise ratio (SNR) of
the measurement of the magnitude of opacity changes and the associated
electron temperature.  The average magnitude of the perturbations is
computed using

\begin{equation}
\langle |\Delta\tau| \rangle \ = \Big[\sum\limits_{i=0}^{M-1} (\overline{{a_i}^2} T_B^{-2}) \ / M \Big]^{1/2}
\end{equation}
and the weighted average electron temperature is computed using

\begin{equation}
\langle T_e \rangle \ = - \sum\limits_{i=0}^{M-1} \overline{ab_i} T_B \ / \sum\limits_{i=0}^{M-1} \overline{{a_i}^2}
\end{equation}
This averaging improves the SNR by approximately $2^{1/2} M^{1/4}$,
where $M$ is the number of difference spectra used in equations (10)
and (11). 

The second method is to take the difference between the spectrum on a
given day from the average of the spectra on all the other days at the
same sidereal time.  In this case the change in opacity and electron
temperature are calculated from equations (4) and (5) and errors in
these quantities can be estimated from the covariance matrix and the
RMS residuals to the weighted least squares fit of the difference
spectrum to function (3). The optimal weighting is equal to
$1/\sigma^2$ where $\sigma$ is the expected noise as a function of
frequency.

\section{Results}

Figures 1 and 2 show the magnitude of the perturbations and the
associated electron temperature derived by the first method.  The
error bars in these figures are computed from the weighted RMS
deviation of the noise-corrected values of $\Delta\tau$ and $T_e$
obtained from equations (6) and (7) from the mean values derived from
equations (10) and (11), divided by $2^{1/2} M^{1/4}$.
The incoherent averaging of results from many difference spectra is needed
to obtain a sufficient SNR for measurements of the electron temperature 
during the night.

Examples of these differences for large perturbations are shown in
Figure 3 along with the least squares fit to these spectra and the
corresponding values of $\Delta\tau$ and electron temperature derived
from equation (4) and (5). In this case the SNR for a single difference
spectrum is sufficient to obtain the electron temperature with the errors
from the least squares analysis of each spectrum. 

The second method allows the variation in ionospheric opacity to be
shown as a function of time for each day as shown in Figure 4.  In
this plot only the points with a good fit to the ionosphere are
plotted. The missing points correspond to data corrupted by solar emission,
RFI, or rain.  Each point is from a 15~minute integration from
independent data.  The high correlation between adjacent points shows
that, in most cases, the time scale for opacity changes in the ionosphere is
longer than 15~minutes and on average is about an hour.

\section{Discussion}

The variation of opacity shown in Figure 1 is similar to the
variations in the radiometric measurements of \textit{Steiger and
  Warwick} [1961] at 18~MHz, who show a plot of attenuation vs local
time on 16 December 1958. Their plot shows variations of ionospheric
attenuation of about 2~dB, which corresponds to 0.7\% at 150~MHz, on 
what they call a "typical day". The plot also shows the attenuation
rising sharply at 6 hours local time and declining at 21 hours in a
manner similar to that shown in Figure 1. The time scale for the
variations is about one hour, consistent with those of Figure~4. 

The measurements of electron temperature shown in Figure~2 indicate
that absorption is occurring at a broad range of altitudes. Figure~5
shows electron temperatures on 2014-04-20 at MRO based on
the MSISE-90 model of \textit{Hedin} [1991] and the IRI 2007 model of
\textit{Bilitza and Reinisch} [2008]. This indicates that the measured
electron temperatures during the day time are comparable with model-based F-region electron temperatures. 

Figure 6 shows absorption of a single 150~MHz ray at 45$^{\circ}$ zenith angle. This also supports
the conclusion that slightly over half of the absorption occurs at
altitudes over 150 km. The shape of the total absorption also agrees
relatively well with Figure 1, which is reasonable, assuming that
variation of opacity is to first order proportional to opacity. The
absorption was computed using a variant of the Appleton-Lassen
equation that ignores the magnetic field [\textit{Hargreaves}, 1969]

\begin{equation}
A = 4.6 \cdot 10^{-5} \int \frac{N_e(h)\nu_e(h)}{\nu_e(h)^2 + \omega^2} dh,
\end{equation}
where $A$ is the attenuation in dB, $N_e(h)$ is altitude 
dependent electron density and $\omega^2$
is the angular frequency of the radio wave. Here, $\nu_e(h)$ is the
electron collision frequency, including not only electron-neutral, but
also electron-ion collisions, which are important in the F-region
[\textit{Mitra and Shain}, 1953]. The electron collision
frequency height distribution $\nu_e = \langle\nu_{\mathrm{ei}}\rangle +
\langle\nu_{\mathrm{en}}\rangle$ function shown in Figure 7 of
\textit{Aggarwal et.al., } [1979] was used to derive the attenuation curves
plotted in Figure~6.

\subsection{Extended Ionosphere model}

The simple model of a single layer, considered in Section 2.1  can be extended to include many layers. In this case, 
if the modeled absorption is small, the spectrum of a ray after passage
through layers on the ionosphere, $T(f)$,  can be written as,

\begin{equation}
T(f) = T_Bf^{-s} - T_B f^{-s} f^{-2} \sum_r L_r + f^{-2} \sum_r L_r T_e(r),
\end{equation}
where $L_r$ is the absorption coefficient for a narrow altitude range interval
at range $r$, $T_Bf^{-s}$, is the spectrum of sky noise above the ionosphere in the direction
from which the ray originates  
and $T_e(r)$ is electron temperature at range $r$. The
sum is formed across the range of all layers though which the ray propagates.
This ignores the change in sky noise spectrum
through the ionosphere owing to
the small losses involved. 

For spectra $T(f)$ and $T'(f)$ taken on different days at the same sidereal time, 
the difference spectrum,  $\Delta T(f)$, is given by

\begin{eqnarray}
\Delta T(f) = (-T_B f^{-s} f^{-2})\sum_r(L_r - L_r') \nonumber \\ 
     + f^{-2}\sum_r (L_r T_e(r)-L'_r T'_e(r)).
\end{eqnarray}
Here the primed variables correspond to changed quantities in the spectrum $T'(f)$.
If the temperatures on the two different days are assumed equal $\tilde{T}_e = \tilde{T}'_e$,
then equation (14) can be simplified to

\begin{equation}
\Delta T(f) = (-T_B f^{-s} f^{-2} + f^{-2} \tilde{T}_e)\Delta\tau,
\end{equation}
In this case for a single ray path  

\begin{equation}
\tilde{T}_e = \left(\sum_r L_r -L'_r  \right)^{-1} \sum_r (L_r - L'_r) T_e(r)
\end{equation}
and

\begin{equation}
\Delta\tau = \sum_r L_r -L'_r 
\end{equation}

If the day to day perturbations in the F-region dominate, the electron temperature
derived from the difference spectrum would be an estimate of the 
electron temperature in the F-region.  
Conversely, if the perturbations in the D-region dominate, the electron temperature
derived from the difference spectrum would be an estimate of the
D-region temperature.

As we currently do not have a way to model the day-to-day variation in
absorption, we have attempted to model the observed electron
temperatures using the approximation that assumes there is a constant
fractional change from day to day in attenuation for each layer, i.e.,
$L_r - L'_r = c L_r$, where $c$ is an arbitrary constant. In this
case, the measured electron temperature is the weighted average of
electron temperature, weighted with attenuation in each layer. This
simplified model, computed for a single $45^{\circ}$ slanted ray going
through the ionosphere is shown in Figure 2, overlayed on the
measurements. We used values of $N_e(r)$ and $T_e(r)$ obtained 
for the MRO location from
the IRI-2007 and MSISE-90 models for our computations.

A more complete modeling of the electron temperature from 
the difference spectra requires estimates of changes in
loss for each layer of the ray path,
horizontal structure, refraction,   
as well as convolution over all the ray paths with the beam of the
dipole antenna used for the measurements. 
  
While the measurements of electron temperature agree reasonably well 
with this model during the
daytime, the night temperatures are consistently lower in the
measurements. The reason for this is unknown, but it
could be caused by the model underestimating the D-region electron
density or overestimating the F-region electron density during the
night. 
 
This study shows a novel method for
measuring the electron temperature using a relatively inexpensive
passive ground based radio remote sensing method. At present the instrument can
only measure the electron temperature of changes in the ionosphere so the results
presented here are weighted by the perturbations in the attenuation through the ionosphere.
As data is acquired over a full year it will be possible to obtain the electron temperature
weighted by the attenuation relative the minimum values observed at each sidereal time but it will
require models or data from other instruments to obtain the electron temperature as a function of altitude.

\section{Conclusions}

The use of a well calibrated broadband spectrometer in the frequency
range 80 to 185 MHz allows the radiometric measurement of the electron
temperature of ionospheric perturbations in addition to the
measurement of the attenuation changes through the ionosphere. The
measurements agree relatively well with theory and previously
published measurements.

\begin{acknowledgments}

This scientific work makes use of the Murchsion Radio-astronomy Observatory, operated by CSIRO.  
We acknowledge the Wajarri Yamatji people as the traditional owners of the Observatory site.   
We thank the MRO Support Facility team, especially Michael Reay and Suzy Jackson, for maintenance of the instrument.
We also acknowledge Dr. Phillip Erickson of MIT Haystack Observatory for helpful discussions on the analysis used in this paper.  
The raw spectral data and the data format can be made available upon
request to Alan Rogers at arogers@haystack.mit.edu.
This work was supported by the U.S. National Science Foundation through research awards for the Experiment to Detect the Global EoR Signature (AST-0905990 and AST-1207761).

\end{acknowledgments}

%% ------------------------------------------------------------------------ %%
%%  REFERENCE LIST AND TEXT CITATIONS
%
% Either type in your references using
% \begin{thebibliography}{}
% \bibitem{}
% Text
% \end{thebibliography}
%

\end{article}

\newpage

\begin{figure}
\centering
\noindent\includegraphics[width=20pc]{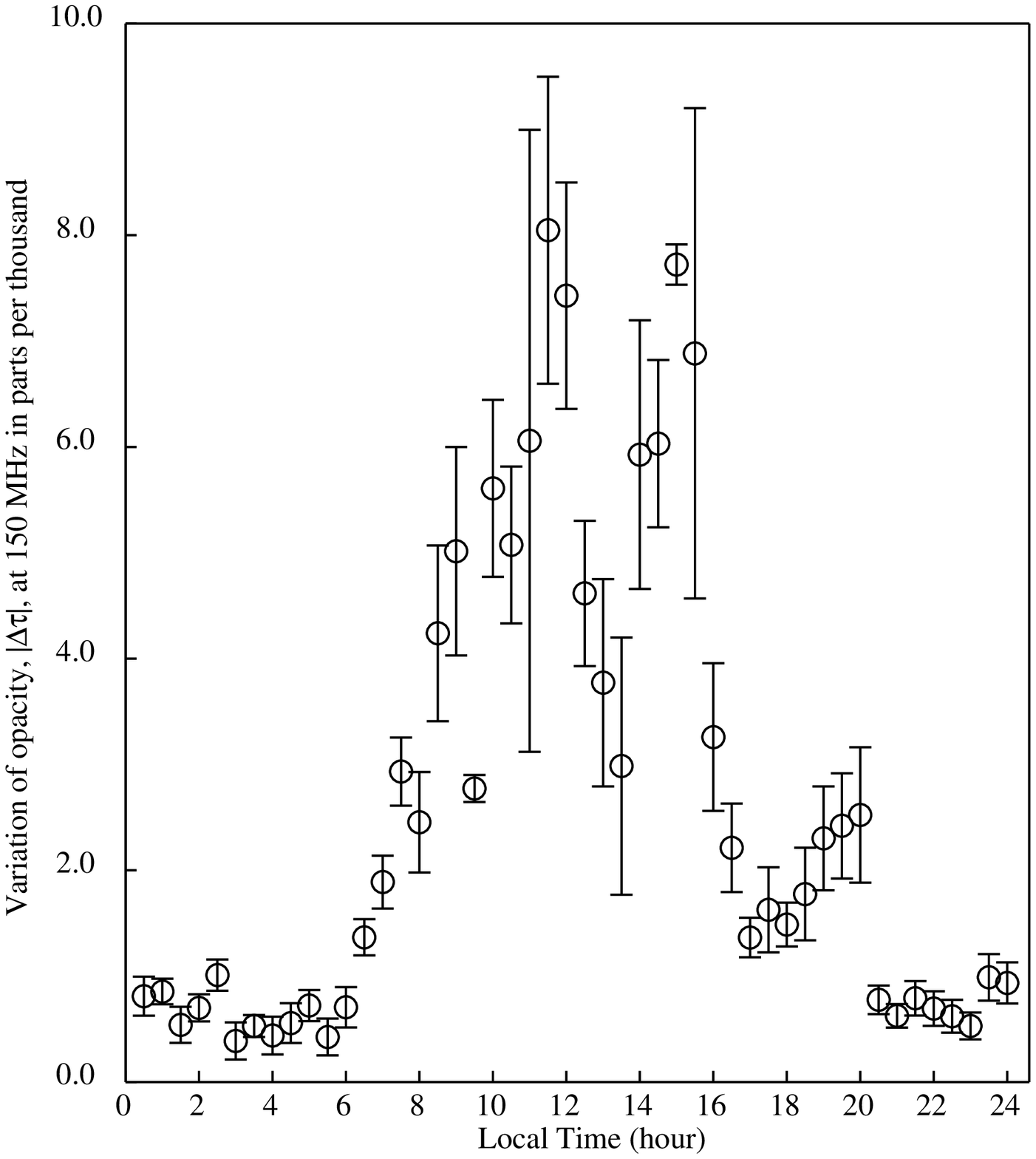}
\caption{Average magnitude of ionospheric perturbations, $\langle |\Delta\tau| \rangle$, vs local time over the 16 days
from 18 April to 6 May 2014. Sunrise and sunset at ground level were at about
6.5 and 17.5 hours localtime respectively. The error bars are $\pm 1 \sigma$}
\label{figure 1}
\end{figure}

\begin{figure}
\centering
\noindent\includegraphics[width=20pc]{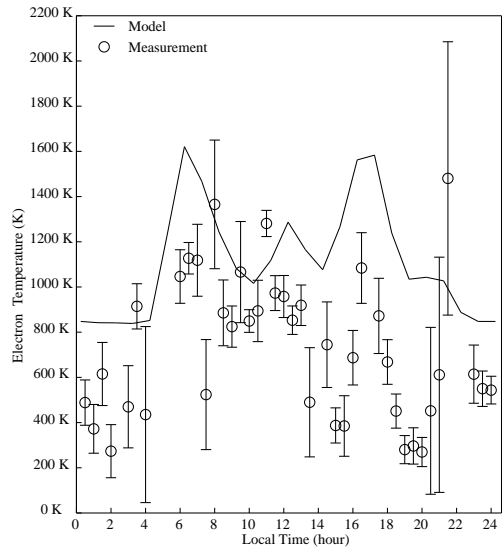}
\caption{Circles represent average electron temperature of the region
  in which perturbations of ionospheric opacity occur vs local
  time. The error bars are $\pm 1 \sigma$. The electron temperature
  was not well determined around 5 hours local time because of a
  combination of small perturbations and large sky noise from the
  transit of the galactic center at this time. The solid line
  represents the weighted sum of electron temperatures obtained from
  the IRI and MSIS model. The sum is weighted with absorption across
  altitudes 60 and 400 km.}
\label{Figure 2}
\end{figure}

\begin{figure}
\centering
\noindent\includegraphics[width=20pc]{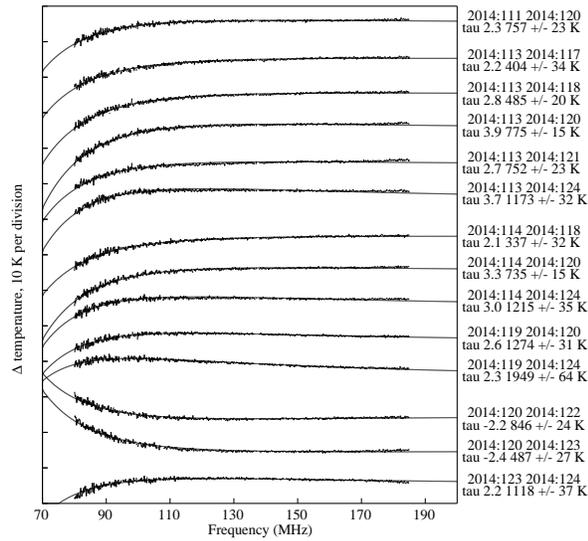}
\caption{Examples of the difference spectrum of the largest
  ionospheric perturbations at the Galactic hour angle of 14 hours
  which corresponds to a local time of about 17.75 hours.
  The dates of the 2 spectra whose difference was taken are given on
  the right of the plot along with the change in opacity at 150 MHz
  in units of parts per thousand
  and the associated electron temperature and 
  $1~\sigma$ error from the best fit
  to the difference spectrum.  An integration of 60 minutes was used
  with a small amount of data excluded due to RFI. The spectra are offset vertically for clarity.}
\label{figure 3}
\end{figure}

\begin{figure}
\centering
\noindent\includegraphics[width=20pc]{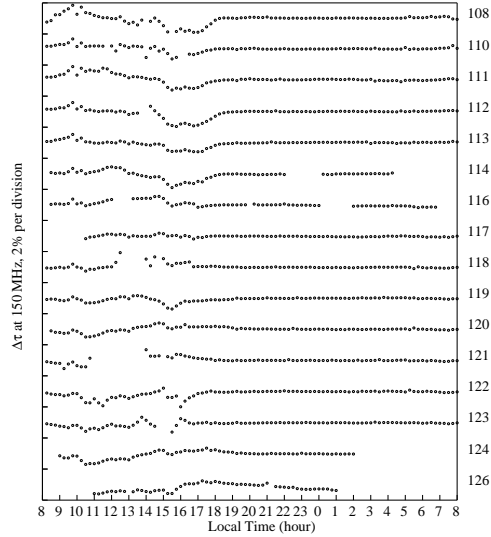}
\caption{Plots of the change of opacity vs local time for each day obtained from the difference between the spectrum 
taken at that local local time an the average spectrum for all days taken at the same sidereal time. The integration time was 15 minutes with some data
excluded owing to RFI and weather conditions.}
\label{figure 4}
\end{figure}

\begin{figure}
\centering
\noindent\includegraphics[width=20pc]{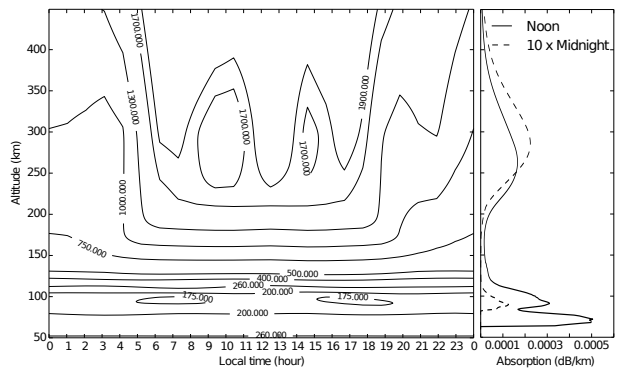}
\caption{Electron temperature and absorption obtained from the IRI ($>$120 km) and MSIS ($<$120 km) models on
on 2014-04-20 for the MRO at $(-26.7^\circ N, 116.6^\circ E)$}
\label{figure 5}
\end{figure}

\begin{figure}
\centering
\noindent\includegraphics[width=20pc]{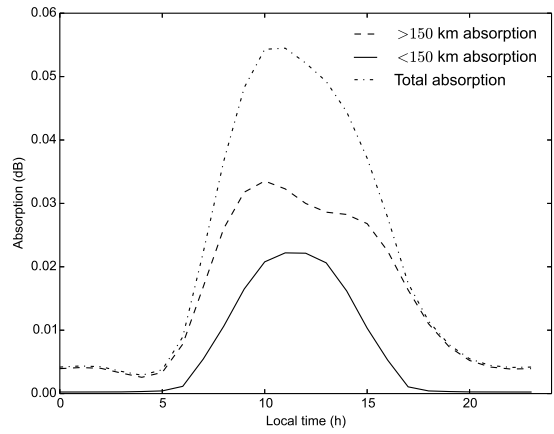}
\caption{Modeled absorption based on electron densities from the
  IRI-2007 model, and collision frequencies based on \textit{Aggarwal
    et.al.} [1979]. The model shows that during normal quiet
  geophysical conditions with normal D-region electron density, the
  F-region has significant absorption.}
\label{figure 6}
\end{figure}

\end{document}